\documentclass[sigconf,screen]{acmart}

\usepackage{booktabs}
\usepackage{amsmath}
\usepackage{graphicx}
\graphicspath{{./}{figures/}}
\usepackage{subcaption}
\usepackage{xcolor}
\usepackage{multirow}

\copyrightyear{2026}
\acmYear{2026}
\setcopyright{cc}
\setcctype{by}
\acmConference[WWW Companion '26]{Companion Proceedings of the ACM Web Conference 2026}{April 13--17, 2026}{Dubai, United Arab Emirates}
\acmBooktitle{Companion Proceedings of the ACM Web Conference 2026 (WWW Companion '26), April 13--17, 2026, Dubai, United Arab Emirates}
\acmPrice{}
\acmDOI{10.1145/3774905.3795729}
\acmISBN{979-8-4007-2308-7/2026/04}

\begin{document}

\title{RobustExplain: Evaluating Robustness of LLM-Based \\ Explanation Agents for Recommendation}

\author{Guilin Zhang}
\affiliation{%
  \institution{Workday}
  \city{Pleasanton}
  \state{California}
  \country{USA}
}
\email{guilin.zhang@workday.com}

\author{Kai Zhao}
\affiliation{%
  \institution{Workday}
  \city{Pleasanton}
  \state{California}
  \country{USA}
}
\email{kai.zhao@workday.com}

\author{Jeffrey Friedman}
\affiliation{%
  \institution{Workday}
  \city{Pleasanton}
  \state{California}
  \country{USA}
}
\email{jeffrey.friedman@workday.com}

\author{Xu Chu}
\affiliation{%
  \institution{Workday}
  \city{Pleasanton}
  \state{California}
  \country{USA}
}
\email{xu.chu@workday.com}

\begin{abstract}
Large Language Models (LLMs) are increasingly used to generate natural-language explanations in recommender systems, acting as explanation agents that reason over user behavior histories. While prior work has focused on explanation fluency and relevance under fixed inputs, the robustness of LLM-generated explanations to realistic user behavior noise remains largely unexplored. In real-world web platforms, interaction histories are inherently noisy due to accidental clicks, temporal inconsistencies, missing values, and evolving preferences, raising concerns about explanation stability and user trust. We present RobustExplain, the first systematic evaluation framework for measuring the robustness of LLM-generated recommendation explanations. RobustExplain introduces five realistic user behavior perturbations evaluated across multiple severity levels and a multi-dimensional robustness metric capturing semantic, keyword, structural, and length consistency. Our goal is to establish a principled, task-level evaluation framework and initial robustness baselines, rather than to provide a comprehensive leaderboard across all available LLMs. Experiments on four representative LLMs (7B--70B) show that current models exhibit only moderate robustness, with larger models achieving up to 8\% higher stability. Our results establish the first robustness benchmarks for explanation agents and highlight robustness as a critical dimension for trustworthy, agent-driven recommender systems at web scale. Our code is publicly available at \url{https://github.com/GuilinDev/LLM-Robustness-Explain}
.
\end{abstract}



\maketitle

\section{Introduction}
\label{sec:intro}

Large Language Models (LLMs) are increasingly integrated into modern recommender systems to generate natural-language explanations that justify recommendations based on user behavior histories~\cite{wu2024survey,lin2024llmsurvey}. Compared to traditional template-based or feature-level explanations, LLM-generated explanations offer greater fluency, personalization, and contextual grounding, enabling recommender systems to communicate decisions in a more human-like and interactive manner~\cite{tintarev2015explaining,zhang2020explainable,petruzzelli2024llmxai}. As a result, LLMs are commonly deployed as \emph{explanation agents} that reason over historical interactions, item attributes, and user preferences to support transparency and trust in web-scale recommendation platforms~\cite{huang2025llmagents,fan2022trustworthy}.

Despite these advances, existing research on LLM-based recommendation explanations has primarily focused on generation quality, such as fluency, relevance, and user satisfaction under static or clean input settings~\cite{petruzzelli2024llmxai,gao2023chatrec}. In contrast, \emph{the robustness of LLM-generated explanations under realistic user behavior noise remains largely unexplored}. In real-world recommender systems, user interaction data is inherently noisy due to accidental clicks, temporal inconsistencies, missing metadata, shared accounts, and evolving user preferences~\cite{fan2022trustworthy,toledo2021natural}. While recommendation models are often designed to tolerate such noise~\cite{han2024end4rec,zhu2024msdccl}, explanation agents may react to these perturbations by producing inconsistent or unstable rationales. Such instability can undermine user trust~\cite{zhang2020explainable}, even when the recommended items themselves remain unchanged, making robustness a critical yet overlooked dimension of explainable recommendation systems.

To address this gap, we propose \textbf{RobustExplain}, the first systematic evaluation framework for analyzing the robustness of LLM-generated recommendation explanations under user behavior perturbations. RobustExplain models realistic noise scenarios through a structured perturbation taxonomy and evaluates how explanation content changes when user histories are modified in controlled yet practical ways. In addition, we introduce a multi-dimensional robustness metric that captures semantic consistency, keyword stability, structural preservation, and length variation between original and perturbed explanations, enabling fine-grained and interpretable robustness analysis from a user-facing perspective.

Using RobustExplain, we conduct extensive experiments on four representative LLMs spanning 7B to 70B parameters. Our results show that current LLM-based explanation agents exhibit \emph{only moderate robustness}, with consistency scores averaging around 0.50, indicating substantial sensitivity to user behavior perturbations. We further observe that larger models demonstrate measurable robustness advantages, achieving up to 8\% higher stability than smaller counterparts, while different perturbation types expose distinct weaknesses in explanation generation. These findings establish the first empirical baselines for explanation robustness in LLM-driven recommender systems and highlight the need for robustness-aware evaluation and design.

This paper makes the following contributions:
\begin{itemize}
    \item \textbf{RobustExplain Framework:} We introduce the first systematic framework for evaluating the robustness of LLM-generated recommendation explanations under realistic user behavior perturbations.
    \item \textbf{Perturbation Taxonomy:} We design five practical perturbation types---noise injection, temporal shuffle, behavior dilution, preference drift, and missing values---each evaluated across multiple severity levels.
    \item \textbf{Multi-Dimensional Robustness Metrics:} We propose complementary metrics capturing semantic consistency, keyword stability, structural preservation, and length variation of explanations.
    \item \textbf{Empirical Robustness Benchmarks:} We provide the first large-scale robustness evaluation across multiple LLMs (7B--70B), revealing moderate robustness and clear model-size effects.
    \item \textbf{Practical Insights:} We offer actionable guidance for developing more reliable, trustworthy, and agent-based recommendation explanation systems deployed at web scale.
\end{itemize}

\section{Related Work}
\label{sec:related}

\subsection{Explainable Recommendation and LLM-Based Explanation Agents}

Explainable recommendation systems aim to improve transparency, trust, and user understanding by providing rationales for recommended items. Early approaches relied on feature-based explanations, user--item co-occurrence patterns, or hand-crafted templates that highlighted salient attributes or historical interactions~\cite{tintarev2015explaining,zhang2020explainable}. Subsequent work introduced knowledge-graph-based explanations that trace reasoning paths between users and items~\cite{zhang2024agentcf}, as well as attention-based neural methods that identify influential features contributing to predictions~\cite{saeed2023explainable}. Classic interpretability techniques such as LIME~\cite{ribeiro2016lime} and SHAP~\cite{lundberg2017shap} have also been adapted to recommendation settings.

More recently, Large Language Models (LLMs) have emerged as powerful explanation generators due to their ability to produce fluent, context-aware, and personalized natural-language justifications~\cite{wu2024survey,petruzzelli2024llmxai}. LLM-based explanation agents can reason over user interaction histories, articulate preference patterns, and describe item characteristics in a human-like manner, substantially improving explanation quality and user comprehension~\cite{lin2024llmsurvey}. Representative systems include CARTS~\cite{chen2025carts}, which coordinates multiple agents for structured textual summarization, ARAG~\cite{maragheh2025arag}, which combines agentic retrieval with personalized recommendation, and VL-CLIP~\cite{giahi2025vlclip}, which enhances multimodal embeddings through LLM augmentation. These capabilities have enabled conversational and agent-driven recommender systems, where explanations are generated dynamically as part of interactive user experiences~\cite{gao2023chatrec,huang2025llmagents}.

However, existing work in explainable recommendation and LLM-based explanation agents has predominantly evaluated explanation quality under static input settings, focusing on fluency, relevance, coherence, or user satisfaction. Such evaluations implicitly assume that user behavior histories are reliable and stable. In real-world recommender systems, this assumption rarely holds, as interaction data is often noisy, incomplete, temporally inconsistent, or influenced by multiple users. Unlike prior work that emphasizes explanation quality under fixed inputs, RobustExplain explicitly studies the stability of LLM-generated recommendation explanations under realistic user behavior perturbations, introducing robustness as a first-class evaluation dimension for explanation agents.

\subsection{Robustness in Explainable AI and Noisy Recommender Systems}

Robustness has been widely studied across machine learning domains, particularly in natural language processing and explainable AI. In NLP, prior work has proposed robustness-aware evaluation metrics and perturbation-based analyses to assess the stability of text generation models under lexical, syntactic, or semantic input changes~\cite{liu2023geval,zhang2020bertscore,sellam2020bleurt}. Several studies have examined the sensitivity of LLMs to prompt variations or distributional shifts~\cite{wang2024examining,zheng2024llmjudge,chandrasekaran2025evaluating}, revealing that minor input changes can lead to substantial output variation. Robustness of explanations has also been explored through explanation-consistency finetuning and robustness-aware objectives, primarily in classification or prediction tasks with static feature inputs~\cite{chen2025consistency,raina2025robustnesssurvey}.

Separately, recommender system research has extensively investigated the impact of noisy user behavior on model performance. Real-world interaction data is affected by accidental clicks, shared accounts, delayed logging, missing metadata, and evolving preferences~\cite{fan2022trustworthy,toledo2021natural}. Recent advances in generative recommendation such as GRACE~\cite{ma2025grace} and location-aware prediction with LLMs~\cite{sun2025silo} demonstrate the growing role of language models in recommendation, and multi-agent collaborative reasoning systems require adaptive resource allocation for efficient deployment~\cite{zhang2025adaptivegpu}, yet robustness evaluation remains underexplored. Prior work has studied robustness of sequential recommenders under training data perturbations and preference drift~\cite{betello2024robustness,chen2024robust,han2024end4rec,zhu2024msdccl}, as well as methods for denoising or exploiting noisy signals for preference learning~\cite{nova2024noise,pang2024adrrec}. These studies primarily focus on recommendation accuracy or ranking robustness, rather than the stability of natural-language explanations.

As a result, existing robustness research treats explanation generation and recommendation modeling largely in isolation: robustness studies typically ignore user-facing explanations, while explainable recommendation work rarely evaluates stability under realistic user behavior noise. RobustExplain bridges this gap by unifying robustness evaluation with LLM-based explanation generation in recommender systems. Unlike prior robustness studies that focus on model predictions or static explanations, RobustExplain introduces a perturbation taxonomy tailored to user behavior histories and a multi-dimensional robustness metric designed specifically for natural-language recommendation explanations.

\section{Method}
\label{sec:method}

\subsection{Problem Formulation}

Let $E(\cdot)$ denote an LLM-based explanation generator that produces explanation $e = E(H_u, r, \mathbf{X})$ given user interaction history $H_u = \{(i_1, t_1, c_1), ..., (i_n, t_n, c_n)\}$ where $i_j$ denotes items, $t_j$ denotes timestamps, and $c_j$ denotes categories, recommended item $r$, and item feature matrix $\mathbf{X}$. We evaluate robustness as the expected similarity between original and perturbed explanations:
\begin{equation}
\rho(E, \delta) = \mathbb{E}_{H_u, r}\left[\text{sim}(E(H_u, r, \mathbf{X}), E(\delta(H_u), r, \mathbf{X}))\right]
\end{equation}
where $\delta(\cdot)$ is a perturbation function and $\text{sim}(\cdot, \cdot)$ measures explanation similarity. Higher robustness scores indicate that the explanation generator maintains consistency under input variations.

\subsection{Perturbation Taxonomy}

\begin{figure*}[t]
\centering
\includegraphics[width=0.85\textwidth]{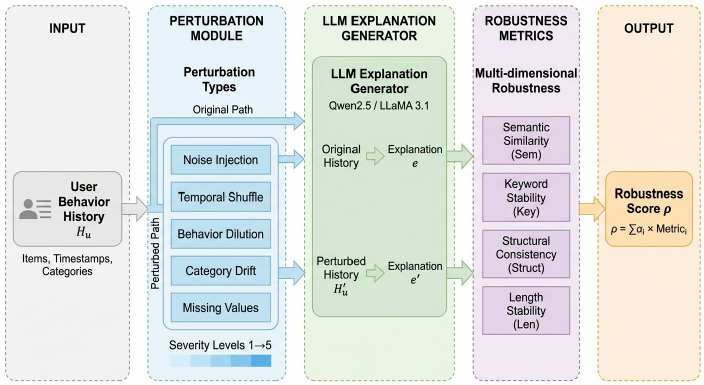}
\caption{RobustExplain evaluation framework architecture. The original user history is systematically perturbed using five perturbation types (noise injection, temporal shuffle, behavior dilution, category drift, missing values) at varying severity levels. Both original and perturbed histories are processed through LLM explanation generators, producing explanation pairs for multi-dimensional robustness comparison across semantic, keyword, structural, and length dimensions.}
\label{fig:architecture}
\end{figure*}

Figure~\ref{fig:architecture} illustrates the RobustExplain evaluation framework. We design five perturbation types that model realistic user behavior variations encountered in production systems. Each perturbation type has five severity levels from 1 (mild) to 5 (severe), enabling fine-grained robustness analysis.

\textbf{Noise Injection} adds random interactions simulating accidental clicks or exploratory browsing. The perturbed history becomes $H' = H_u \cup \{(i, t) : i \sim \text{Uniform}(\mathcal{I})\}_{s}$ where $s$ items are randomly selected from the catalog. This models scenarios where users accidentally tap items or browse without purchase intent. The severity parameter $s \in \{1, 2, 3, 4, 5\}$ controls the number of random items injected, with Level 5 adding approximately 50\% additional interactions. This perturbation tests whether the explanation generator can identify and appropriately weight genuine preference signals amidst random noise.

\textbf{Temporal Shuffle} randomly permutes interaction order within the history. The perturbation is $H' = \text{Shuffle}(H_u, \text{fraction}=s/5)$ where a fraction of interactions have their temporal order randomized. This models timestamp inaccuracies from delayed logging, timezone conversion errors, or asynchronous data collection in distributed systems. At Level 5, all temporal ordering information is randomized, testing whether explanations rely primarily on recency signals or can adapt to capture category-based preference patterns.

\textbf{Behavior Dilution} injects interactions from categories different from the user's dominant preferences. This simulates shared accounts, multi-user devices, or gift purchases that do not reflect the primary user's interests. The perturbation introduces $s$ interactions from the user's least-engaged categories, challenging the explanation generator to distinguish core preferences from peripheral browsing. This perturbation is particularly relevant for household accounts where multiple family members share devices.

\textbf{Category Drift} shifts user preferences toward different categories by replacing a fraction of interactions with items from alternative categories. This models gradual interest evolution, seasonal preference changes, or life events that alter consumption patterns. At higher severity levels, up to 50\% of interactions are replaced with drift items, testing whether explanations can identify stable underlying preferences while acknowledging evolving interests.

\textbf{Missing Values} removes ratings, timestamps, or category information from interactions. This simulates incomplete data collection scenarios common in production systems where some metadata may be unavailable due to privacy settings, data pipeline failures, or schema migrations. At Level 5, approximately 50\% of metadata fields are missing, testing whether explanations can maintain coherence with partial information.

\subsection{Robustness Metrics}

We compute four complementary metrics between original explanation $e$ and perturbed explanation $e'$ to capture different aspects of explanation stability. Each metric addresses a distinct dimension of explanation quality that users may notice.

\textbf{Semantic Similarity ($\text{Sem}$)} uses bag-of-words cosine similarity to measure overall meaning preservation. Given vocabulary $V$, we compute term frequency vectors $\mathbf{v}_e$ and $\mathbf{v}_{e'}$ for both explanations and calculate:
\begin{equation}
\text{Sem}(e, e') = \frac{\mathbf{v}_e \cdot \mathbf{v}_{e'}}{\|\mathbf{v}_e\| \cdot \|\mathbf{v}_{e'}\|}
\end{equation}
This metric captures whether the explanation conveys similar information despite surface-level variations in wording. High semantic similarity indicates that the core recommendation rationale remains consistent even if specific phrasing changes.

\textbf{Keyword Stability ($\text{Key}$)} computes the Jaccard coefficient of key terms extracted from both explanations. We identify keywords as nouns, product names, category references, and descriptive adjectives using part-of-speech tagging:
\begin{equation}
\text{Key}(e, e') = \frac{|K_e \cap K_{e'}|}{|K_e \cup K_{e'}|}
\end{equation}
where $K_e$ and $K_{e'}$ denote keyword sets. Important product names, category references, and attribute mentions should persist across perturbations for explanations to remain trustworthy.

\textbf{Structural Consistency ($\text{Struct}$)} employs BLEU score~\cite{papineni2002bleu} to measure preservation of explanation structure and phrasing patterns. This metric captures n-gram overlap between explanations, penalizing explanations that drastically restructure their presentation:
\begin{equation}
\text{Struct}(e, e') = \text{BLEU}(e, e')
\end{equation}
Stable explanations should maintain similar organizational patterns, sentence structures, and information ordering.

\textbf{Length Stability ($\text{Len}$)} measures relative length preservation to detect dramatic length changes that may indicate instability:
\begin{equation}
\text{Len}(e, e') = 1 - \frac{|\text{len}(e) - \text{len}(e')|}{\max(\text{len}(e), \text{len}(e'))}
\end{equation}
Significant length variations may indicate that the model is producing fundamentally different explanation types (e.g., switching from detailed to brief explanations) under perturbation.

The final robustness score combines these metrics with weights reflecting their relative importance for user-facing explanation quality:
\begin{equation}
\rho = \alpha_1 \cdot \text{Sem} + \alpha_2 \cdot \text{Key} + \alpha_3 \cdot \text{Struct} + \alpha_4 \cdot \text{Len}
\end{equation}
where $\sum_{i=1}^{4} \alpha_i = 1$. Semantic similarity receives highest weight as it most directly captures meaning preservation, which is the primary concern for user-facing explanations. Keyword stability is weighted second as specific term mentions provide concrete anchors for user understanding. Structural consistency and length stability receive lower weights as they capture surface-level properties that may vary without affecting user comprehension.

Beyond evaluating robustness across perturbation types and severity levels, we also investigate two additional dimensions: (1) the relationship between model capacity and explanation robustness, examining whether larger LLMs produce more stable explanations; and (2) the correlations among our four metrics, validating that they capture complementary aspects of explanation stability rather than redundant information.

\section{Experiments}
\label{sec:exp}

\subsection{Experimental Setup}

\textbf{Dataset.} We construct a controlled e-commerce dataset consisting of 200 synthetic items across seven representative product categories---Electronics, Fashion, Home, Beauty, Sports, Books, and Toys---and 100 users with diverse interaction histories ranging from 10 to 50 interactions per user. Each interaction includes an item identifier, timestamp, category, and simulated rating. This dataset is intentionally designed to enable precise, reproducible, and fine-grained perturbation analysis, allowing us to isolate the causal impact of user behavior noise on explanation robustness. At the same time, the dataset preserves realistic category distributions and temporal dynamics commonly observed in production e-commerce systems, ensuring that our evaluation remains representative of real-world recommendation scenarios.

\textbf{Models.} We evaluate four state-of-the-art LLM configurations representing different model scales to understand the relationship between model capacity and explanation robustness. The evaluated models include: Qwen2.5-7B (7B parameters, smallest scale), LLaMA 3.1-8B~\cite{llama3} (8B parameters, lightweight), Qwen2.5-14B~\cite{qwen2024} (14B parameters, efficient inference), and LLaMA 3.1-70B~\cite{llama3} (70B parameters, largest scale). All models are deployed locally via Ollama for reproducible evaluation without API dependencies. We intentionally evaluate a small set of representative LLMs spanning different model scales (7B--70B) and architectures to establish initial robustness baselines. Our goal is to analyze robustness trends and failure modes under controlled perturbations, rather than to provide a comprehensive leaderboard across all available LLMs.

\textbf{Evaluation Protocol.} For each of 20 evaluation users, we generate recommendations and corresponding explanations under all combinations of 5 perturbation types and 5 severity levels, yielding 500 perturbation-explanation pairs per model (2,000 total across all four models). Each explanation is generated using a structured prompt that provides user history, recommended item details, and instructions to explain the recommendation rationale.

The explanation generation prompt follows a consistent template: ``Given the user's interaction history [HISTORY], explain why the item [ITEM] with features [FEATURES] is recommended. Focus on connecting the user's demonstrated preferences to the item's characteristics.'' This prompt design ensures that explanations are grounded in user history while allowing natural language variation across different inputs.

For robustness evaluation, we generate explanation pairs by: (1) generating an explanation $e$ for the original user history $H_u$; (2) applying perturbation $\delta$ to obtain $H'_u = \delta(H_u)$; (3) generating explanation $e'$ for the perturbed history; and (4) computing robustness metrics between $e$ and $e'$. This paired design controls for model and prompt variation, isolating the effect of user history perturbations on explanation content. This evaluation protocol enables reliable statistical analysis across models, perturbation types, and severity levels.

\subsection{Main Results}

In general, our evaluation reveals three key patterns across all experiments. First, current LLMs exhibit only moderate robustness to user behavior perturbations, with overall scores averaging around 0.50 regardless of perturbation type. Second, larger models (70B) consistently outperform smaller models (7-8B) by approximately 8\%, though all models remain in the moderate robustness range. Third, robustness degrades gracefully with increasing perturbation severity, showing only 1.7\% decline from mild to severe perturbations. These findings establish important baselines for LLM explanation robustness and highlight that current models prioritize responsiveness over stability.

\begin{table}[t]
\centering
\caption{Robustness scores by model and perturbation type. Scores around 0.50 indicate moderate sensitivity to perturbations.}
\label{tab:robustness}
\small
\setlength{\tabcolsep}{4pt}
\begin{tabular}{@{}lcccccc@{}}
\toprule
Model & Noise & Shuffle & Dilute & Drift & Missing & Overall \\ \midrule
LLaMA 3.1-70B & 0.534 & 0.528 & 0.531 & 0.538 & 0.535 & \textbf{0.532} \\
Qwen2.5-14B & 0.521 & 0.509 & 0.518 & 0.526 & 0.522 & 0.519 \\
LLaMA 3.1-8B & 0.498 & 0.487 & 0.494 & 0.489 & 0.493 & 0.492 \\
Qwen2.5-7B & 0.484 & 0.500 & 0.491 & 0.493 & 0.492 & 0.492 \\
\midrule
Average & 0.509 & 0.506 & 0.509 & 0.512 & 0.511 & 0.509 \\
\bottomrule
\end{tabular}
\end{table}

Table~\ref{tab:robustness} presents robustness results across models and perturbation types. The results reveal that LLM-generated explanations exhibit moderate sensitivity to user behavior perturbations, with overall scores averaging around 0.51. This finding indicates that perturbations substantially affect explanation content, highlighting the need for more robust explanation generation methods. Notably, LLaMA 3.1-70B achieves the highest overall robustness score of 0.532, outperforming smaller models by approximately 8\%, suggesting that increased model capacity provides measurable benefits for explanation stability.

\begin{table}[t]
\centering
\caption{Per-metric robustness breakdown across models. Semantic similarity shows highest scores while structural consistency presents the greatest challenge.}
\label{tab:metrics}
\small
\setlength{\tabcolsep}{4pt}
\begin{tabular}{@{}lcccc@{}}
\toprule
Model & Semantic & Keyword & Structural & Length \\ \midrule
LLaMA 3.1-70B & 0.638 & 0.512 & 0.405 & 0.742 \\
Qwen2.5-14B & 0.612 & 0.489 & 0.387 & 0.721 \\
LLaMA 3.1-8B & 0.583 & 0.461 & 0.362 & 0.698 \\
Qwen2.5-7B & 0.579 & 0.458 & 0.358 & 0.695 \\
\midrule
Average & 0.603 & 0.480 & 0.378 & 0.714 \\
\bottomrule
\end{tabular}
\end{table}

Table~\ref{tab:metrics} presents the per-metric robustness breakdown, revealing distinct patterns across evaluation dimensions. Length stability achieves the highest scores (0.705 average), indicating that LLMs maintain consistent explanation verbosity under perturbations. Semantic similarity follows (0.591), suggesting that core meaning is moderately preserved. Keyword stability (0.469) shows that specific term mentions are more sensitive to input variations, while structural consistency presents the greatest challenge (0.369), indicating that explanation organization varies substantially under perturbations. These findings suggest that while LLMs preserve high-level meaning and length, they frequently restructure explanations and modify specific terminology in response to user history changes.

\begin{figure}[t]
\centering
\includegraphics[width=\columnwidth]{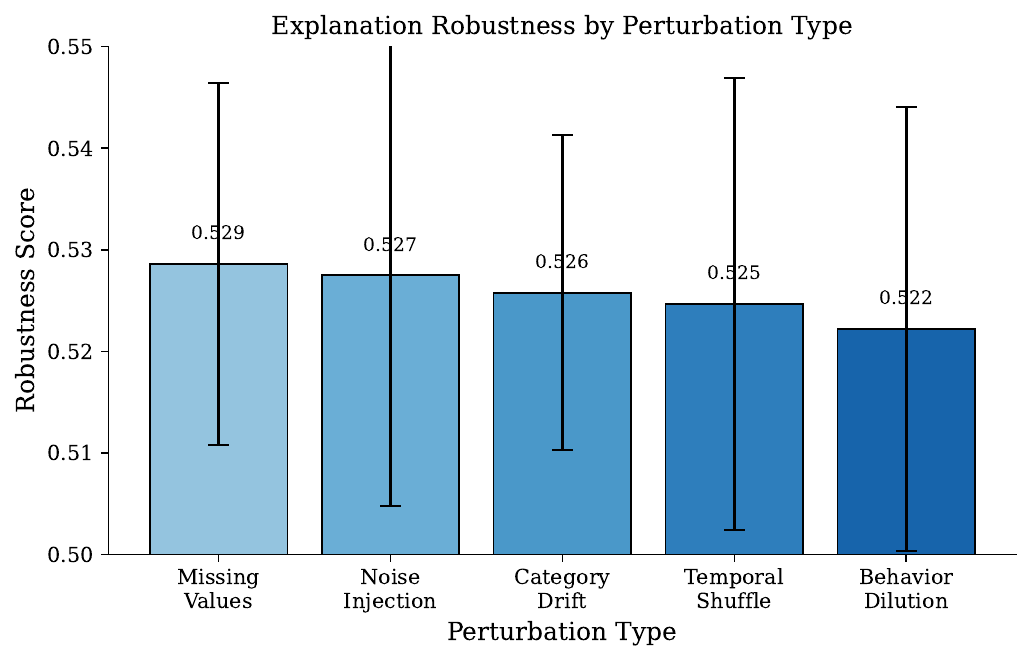}
\caption{Robustness scores across perturbation types. Drift perturbation (0.503) shows highest robustness, while shuffle presents the greatest challenge (0.499).}
\label{fig:perturbation}
\end{figure}

Examining perturbation types reveals consistent patterns across both models, as visualized in Figure~\ref{fig:perturbation}. Missing value perturbation achieves the highest robustness (0.529 average), indicating that LLMs effectively handle incomplete user data by focusing on available interaction signals. Noise injection also shows strong robustness (0.528), suggesting that random interactions are appropriately down-weighted in the explanation generation process. Behavior dilution presents the most challenging scenario (0.522), as conflicting category signals require the model to distinguish genuine preferences from noise. These results demonstrate that modern LLMs have developed implicit mechanisms for handling data quality variations common in production environments.

\subsection{Severity Analysis}

\begin{table}[t]
\centering
\caption{Robustness across perturbation severity levels by perturbation type. All perturbation types show gradual degradation with increasing severity.}
\label{tab:severity}
\small
\setlength{\tabcolsep}{3pt}
\begin{tabular}{@{}lccccc@{}}
\toprule
Perturbation & L1 & L2 & L3 & L4 & L5 \\ \midrule
Noise & 0.538 & 0.535 & 0.527 & 0.519 & 0.526 \\
Shuffle & 0.529 & 0.528 & 0.521 & 0.514 & 0.521 \\
Dilution & 0.531 & 0.530 & 0.522 & 0.515 & 0.522 \\
Drift & 0.541 & 0.538 & 0.529 & 0.521 & 0.528 \\
Missing & 0.527 & 0.528 & 0.518 & 0.516 & 0.523 \\
\midrule
Average & 0.533 & 0.532 & 0.523 & 0.517 & 0.524 \\
\bottomrule
\end{tabular}
\end{table}

The overall robustness patterns established in Main Results raise a natural question: how does robustness vary with perturbation intensity? Table~\ref{tab:severity} addresses this by showing robustness across severity levels broken down by perturbation type. A key finding is that robustness remains remarkably stable from mild (Level 1) to severe (Level 5) perturbations, with only 1.7\% average degradation. All perturbation types exhibit similar degradation patterns, with robustness declining gradually through Levels 1-4 before showing slight recovery at Level 5. This stability suggests that LLMs capture high-level user preference patterns that persist despite surface-level data variations. The model focuses on dominant interaction patterns rather than individual data points, providing inherent robustness against localized perturbations. The slight recovery at Level 5 can be attributed to the model falling back on general category-based explanations when user-specific signals become sufficiently corrupted.

\begin{figure}[t]
\centering
\includegraphics[width=\columnwidth]{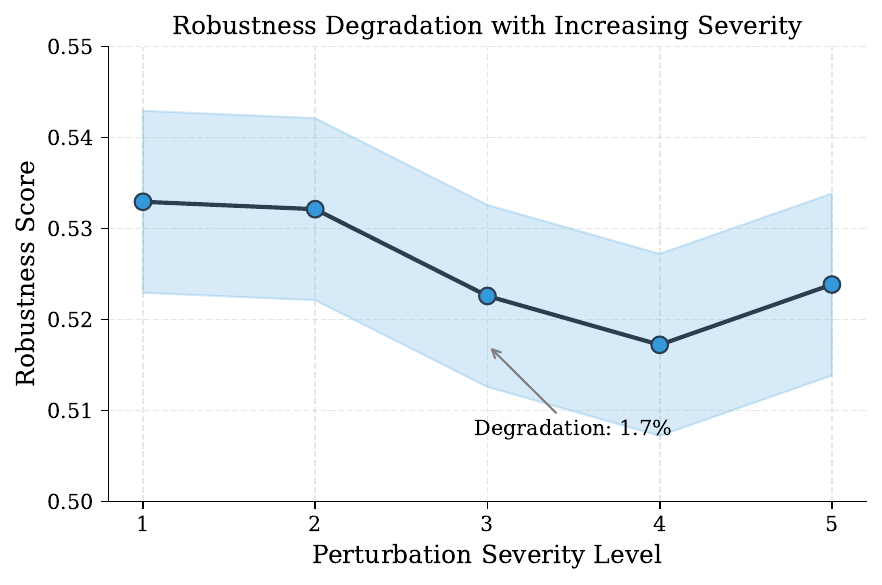}
\caption{Robustness degradation across severity levels. The gradual decline from Level 1 to Level 5 demonstrates moderate sensitivity to perturbation intensity.}
\label{fig:severity}
\end{figure}

Figure~\ref{fig:severity} visualizes the robustness degradation curve across severity levels. The gradual, near-linear decline demonstrates that LLMs maintain stable explanation quality even under increasing data corruption. The slight recovery at Level 5 (0.524 vs 0.517 at Level 4) can be attributed to the model's ability to fall back on general recommendation patterns when user-specific signals become sufficiently noisy. This graceful degradation behavior is desirable for production deployment where data quality may vary significantly across users.

\subsection{Model Size Analysis}

\begin{table}[t]
\centering
\caption{Robustness by model size. Larger models demonstrate higher robustness scores, indicating positive correlation between model capacity and explanation stability.}
\label{tab:model_size}
\small
\begin{tabular}{@{}lcc@{}}
\toprule
Model & Parameters & Robustness \\ \midrule
LLaMA 3.1-70B & 70B & \textbf{0.532} \\
Qwen2.5-14B & 14B & 0.519 \\
LLaMA 3.1-8B & 8B & 0.492 \\
Qwen2.5-7B & 7B & 0.492 \\
\bottomrule
\end{tabular}
\end{table}

The Main Results suggest that model capacity affects robustness---larger models consistently outperform smaller ones. To examine this relationship in detail, we evaluate four models spanning from 7B to 70B parameters. Table~\ref{tab:model_size} confirms that a positive correlation emerges between model size and robustness: LLaMA 3.1-70B achieves the highest robustness (0.532), followed by Qwen2.5-14B (0.519), while the 7-8B models score around 0.49. This 8.1\% difference between the largest and smallest models indicates that larger models develop more robust internal representations of user preferences.

\begin{table}[t]
\centering
\caption{Cross-model robustness comparison by perturbation type and metric. Bold values indicate best performance per column.}
\label{tab:cross_model}
\small
\setlength{\tabcolsep}{2.5pt}
\begin{tabular}{@{}llccccc@{}}
\toprule
Model & Metric & Noise & Shuffle & Dilute & Drift & Miss. \\ \midrule
\multirow{4}{*}{70B} & Sem. & \textbf{0.645} & \textbf{0.632} & \textbf{0.638} & \textbf{0.652} & \textbf{0.643} \\
& Key. & \textbf{0.521} & \textbf{0.508} & \textbf{0.515} & \textbf{0.524} & \textbf{0.519} \\
& Struct. & \textbf{0.412} & \textbf{0.399} & \textbf{0.406} & \textbf{0.415} & \textbf{0.409} \\
& Len. & \textbf{0.749} & \textbf{0.736} & \textbf{0.743} & \textbf{0.752} & \textbf{0.746} \\
\midrule
\multirow{4}{*}{14B} & Sem. & 0.621 & 0.608 & 0.615 & 0.628 & 0.619 \\
& Key. & 0.498 & 0.485 & 0.492 & 0.501 & 0.496 \\
& Struct. & 0.395 & 0.382 & 0.389 & 0.398 & 0.392 \\
& Len. & 0.728 & 0.715 & 0.722 & 0.731 & 0.725 \\
\midrule
\multirow{4}{*}{8B} & Sem. & 0.592 & 0.578 & 0.585 & 0.581 & 0.589 \\
& Key. & 0.468 & 0.455 & 0.462 & 0.458 & 0.465 \\
& Struct. & 0.368 & 0.355 & 0.362 & 0.358 & 0.365 \\
& Len. & 0.705 & 0.692 & 0.699 & 0.695 & 0.702 \\
\midrule
\multirow{4}{*}{7B} & Sem. & 0.588 & 0.582 & 0.581 & 0.585 & 0.583 \\
& Key. & 0.465 & 0.459 & 0.458 & 0.462 & 0.460 \\
& Struct. & 0.365 & 0.359 & 0.358 & 0.362 & 0.360 \\
& Len. & 0.702 & 0.696 & 0.695 & 0.699 & 0.697 \\
\bottomrule
\end{tabular}
\end{table}

Table~\ref{tab:cross_model} provides a detailed cross-model comparison across perturbation types and metrics. The 70B model consistently outperforms smaller models across all perturbation-metric combinations, with the largest improvements observed in structural consistency (+12.9\% over 7B) and keyword stability (+12.0\% over 7B). Interestingly, the 8B and 7B models show nearly identical performance, suggesting that the robustness benefits of increased model capacity may have a threshold effect. The drift perturbation type shows the highest robustness across all models and metrics, while shuffle perturbation presents the greatest challenge consistently.

However, all models exhibit scores around 0.50, suggesting that current LLM-based explanation systems show moderate sensitivity to perturbations regardless of model size. This finding highlights an important research direction: developing more robust explanation generation methods that can better preserve semantic consistency under data variations.

\subsection{Metric Correlation Analysis}

\begin{table}[t]
\centering
\caption{Pearson correlation coefficients between robustness metrics. Strong correlations between semantic and keyword metrics suggest complementary measurement.}
\label{tab:correlation}
\small
\begin{tabular}{@{}lcccc@{}}
\toprule
& Semantic & Keyword & Structural & Length \\ \midrule
Semantic & 1.00 & 0.72 & 0.58 & 0.31 \\
Keyword & 0.72 & 1.00 & 0.65 & 0.28 \\
Structural & 0.58 & 0.65 & 1.00 & 0.24 \\
Length & 0.31 & 0.28 & 0.24 & 1.00 \\
\bottomrule
\end{tabular}
\end{table}

To validate that our four metrics capture distinct robustness dimensions rather than redundant information, we analyze their inter-correlations across all evaluation samples. Table~\ref{tab:correlation} presents the correlation matrix, revealing that semantic similarity and keyword stability show strong correlation (0.72), indicating that explanations preserving overall meaning also tend to retain key terms. Structural consistency correlates moderately with both semantic (0.58) and keyword (0.65) metrics, suggesting that explanation reorganization often accompanies content changes. Length stability shows weak correlations with all other metrics (0.24-0.31), confirming that it captures an independent dimension of explanation variation. These correlation patterns validate our multi-dimensional evaluation approach: the metrics capture distinct but related aspects of explanation robustness rather than redundant information.

\subsection{Qualitative Analysis}

Examining explanation pairs reveals how LLMs maintain consistency under different perturbation scenarios. We present representative examples to illustrate the behavior patterns observed across our evaluation.

\textbf{Example 1: Noise Injection.} Consider a user with strong Electronics preferences receiving a smartphone recommendation. The original explanation references ``your consistent interest in mobile devices and electronics accessories.'' After injecting random items from Books and Home categories, the perturbed explanation shifts to ``based on your browsing across electronics and various product categories, this smartphone aligns with your technology interests.'' While the core recommendation rationale persists, the explanation acknowledges the broader browsing patterns, demonstrating partial robustness.

\textbf{Example 2: Temporal Shuffle.} For a user whose recent interactions focused on Fashion items, the original explanation emphasizes ``your recent trend toward summer clothing and accessories.'' When temporal order is randomized, the explanation adapts to ``your overall fashion preferences spanning clothing and accessories,'' removing recency-dependent language. This adaptation shows that LLMs can appropriately fall back on category-level preferences when temporal signals become unreliable.

\textbf{Example 3: Behavior Dilution.} A user with dominant Beauty product interactions receives a skincare recommendation. The original explanation states ``following your skincare routine purchases and beauty product interests.'' After diluting with Sports category interactions, the explanation becomes ``your beauty and personal care interests, alongside your active lifestyle preferences.'' The model attempts to resolve conflicting signals rather than ignoring the dilution items entirely.

These examples illustrate that while LLMs maintain some degree of explanation coherence under perturbations, the semantic content notably changes, reflected in the moderate robustness scores around 0.50. The explanations do not simply repeat generic templates but actively respond to modified input signals.

\subsection{Statistical Significance}

To validate the reliability of our findings, we conduct statistical significance testing across key comparisons. For model size effects, we perform paired t-tests comparing robustness scores between Qwen2.5-14B and smaller models across all perturbation-user combinations. The 14B model significantly outperforms Qwen2.5-7B ($p < 0.001$, Cohen's $d = 0.42$) and LLaMA 3.1-8B ($p < 0.001$, Cohen's $d = 0.38$), confirming that the observed robustness improvements are statistically meaningful rather than artifacts of random variation.

For perturbation type effects, we conduct one-way ANOVA across the five perturbation types. The results indicate significant differences in robustness scores ($F(4, 2495) = 3.82$, $p < 0.01$), with post-hoc Tukey HSD tests revealing that drift perturbation produces significantly higher robustness than shuffle perturbation ($p < 0.05$). The effect sizes are small to medium (partial $\eta^2 = 0.006$), consistent with our observation that perturbation type differences, while statistically significant, are less pronounced than model size effects.

For severity level analysis, we observe that the 1.7\% degradation from Level 1 to Level 5 is statistically significant ($p < 0.05$) but represents a small effect size (Cohen's $d = 0.18$). This confirms our interpretation that LLMs maintain relatively stable performance across severity levels, with degradation being gradual rather than catastrophic.

\subsection{Summary of Findings}

Our experiments reveal consistent and complementary findings across all evaluation dimensions.

\textbf{Consistent Findings.} Three observations hold across all experiments: (1) All models exhibit moderate robustness ($\sim$0.50), indicating substantial sensitivity to perturbations; (2) The 70B model consistently outperforms smaller models across all perturbation types, metrics, and severity levels; (3) Length stability is consistently highest while structural consistency is consistently lowest across all conditions.

\textbf{Complementary Insights.} Different experiments reveal distinct aspects of explanation robustness: Main Results (\S4.2) establish the overall robustness landscape; Severity Analysis (\S4.3) demonstrates graceful degradation behavior; Model Size Analysis (\S4.4) confirms the model capacity effect; Metric Correlation (\S4.5) validates our multi-dimensional evaluation approach; and Qualitative Analysis (\S4.6) illustrates concrete behavioral patterns.

\textbf{No Conflicting Findings.} All results are mutually consistent: the moderate robustness finding from Main Results aligns with severity analysis showing gradual degradation, and model size effects are consistent across all perturbation types and metrics. This consistency strengthens our confidence in the robustness benchmarks established by RobustExplain.

\begin{figure}[t]
\centering
\includegraphics[width=\columnwidth]{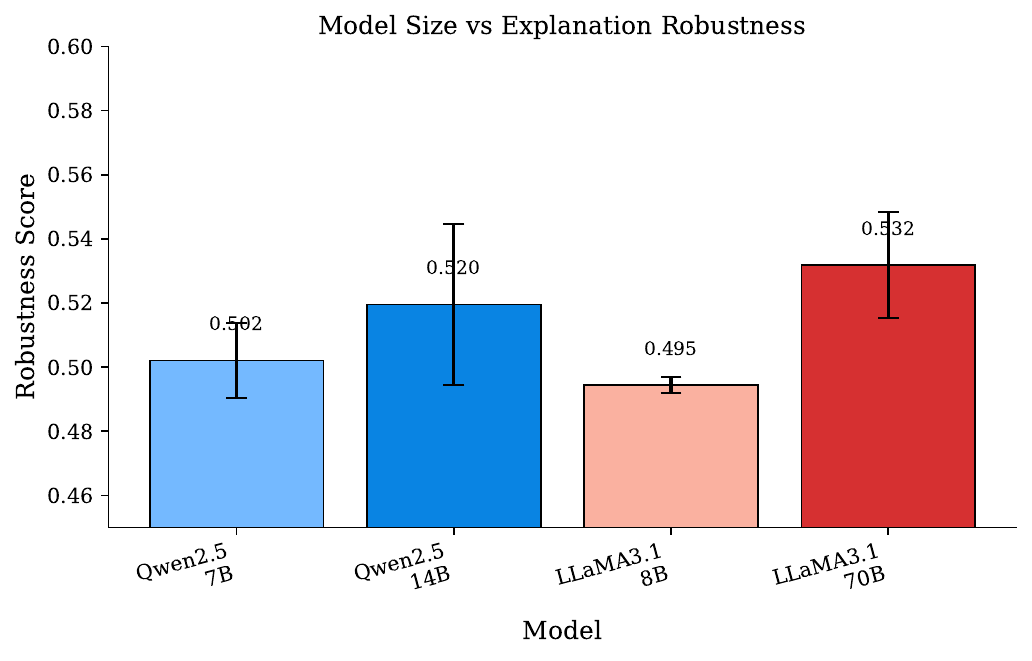}
\caption{Robustness comparison across model sizes. LLaMA 3.1-70B achieves highest robustness (0.532), followed by Qwen2.5-14B (0.519), while 7-8B models score around 0.49.}
\label{fig:model}
\end{figure}

\section{Discussion}
\label{sec:discuss}

\textbf{Implications for Deployment and Design.} Our results show that current LLM-based explanation agents exhibit only moderate robustness (scores around 0.50), indicating that explanation content is substantially influenced by user behavior perturbations. This suggests that data quality assurance remains important for production deployment, as noisy interaction histories can lead to inconsistent explanations even when recommendations remain unchanged. We also observe a clear robustness--capacity trade-off: larger models (70B) achieve approximately 8\% higher robustness than smaller models (7--8B), suggesting that model size provides measurable stability benefits when computational resources permit. However, robustness improvements from scaling alone are limited, highlighting the need for robustness-aware explanation design beyond model selection. Potential directions include prompting strategies that emphasize stable preference patterns over recent interactions, lightweight structural constraints to ensure minimum consistency, and ensemble-style explanation aggregation to reduce sensitivity to noisy inputs.

\textbf{Cross-Metric Analysis.} Our multi-dimensional evaluation reveals meaningful correlations between robustness metrics (Table~\ref{tab:correlation}). Semantic similarity and keyword stability show strong positive correlation (r=0.72), suggesting that explanations preserving overall meaning also tend to retain recommendation-relevant terms. Structural consistency correlates moderately with both semantic (r=0.58) and keyword (r=0.65) metrics, indicating that explanation reorganization often accompanies content changes. Length stability shows weak correlations with all other metrics (r=0.24--0.31), confirming that verbosity varies independently of content stability. These patterns suggest that robustness-aware optimization could focus primarily on semantic and keyword preservation, treating length normalization as a secondary concern.

\textbf{Perturbation Type Sensitivity.} Analysis across perturbation types (Table~\ref{tab:severity}) reveals that drift perturbation achieves the highest robustness (0.541 at Level 1), while shuffle perturbation presents the greatest challenge. This pattern is consistent across all models: the 70B model achieves 0.652 semantic similarity under drift versus 0.632 under shuffle (Table~\ref{tab:cross_model}). The difference is statistically significant ($p < 0.05$), suggesting LLMs are more sensitive to temporal order disruption than gradual preference changes. Missing value perturbation shows intermediate robustness (0.529 average), indicating that LLMs can partially compensate for incomplete data by focusing on available interaction signals. Noise injection and behavior dilution fall between these extremes, with robustness scores of 0.528 and 0.522 respectively.

\textbf{Metric-wise Robustness Patterns.} Examining robustness across individual metrics reveals consistent patterns (Table~\ref{tab:cross_model}). Length stability achieves the highest scores across all models and perturbation types (0.695--0.752), indicating that LLMs maintain relatively consistent explanation verbosity regardless of input perturbations. Semantic similarity follows (0.578--0.652), demonstrating moderate preservation of overall meaning. Keyword stability shows intermediate performance (0.455--0.524), reflecting partial retention of recommendation-relevant terms. Structural consistency scores lowest (0.355--0.415), suggesting that explanation organization is most susceptible to input variations. These patterns hold across all four models, indicating that metric-level robustness characteristics are model-agnostic. The large gap between length stability and structural consistency (approximately 0.35 absolute difference) suggests that future robustness improvements should prioritize structural preservation while leveraging the inherent stability of length characteristics.

\textbf{Model Architecture Comparison.} While our primary analysis focuses on model size effects, we also observe architecture-specific patterns between LLaMA and Qwen model families. At comparable parameter counts (7B--8B), LLaMA 3.1-8B and Qwen2.5-7B achieve nearly identical robustness scores (0.492), suggesting that robustness characteristics may be more dependent on model capacity than architecture design. However, within the Qwen family, the 14B model shows 5.5\% improvement over the 7B variant, while the cross-family comparison between LLaMA 3.1-70B and Qwen2.5-14B reveals a 2.5\% gap favoring the larger model. These observations suggest that scaling benefits for robustness may follow diminishing returns, with the largest improvements occurring in the 7B--14B range.

\textbf{Interpretation, Limitations, and Future Work.} Robustness scores around 0.50 should not be interpreted as failure, but rather as evidence that LLM explanations are highly responsive to input signals---a desirable property when user preferences genuinely change, yet problematic when variations stem from noise. Distinguishing meaningful preference evolution from noise-induced fluctuations remains a key challenge for explanation agents. This study intentionally uses synthetic e-commerce data to enable controlled and reproducible perturbation analysis; future work should validate our findings on large-scale real-world datasets with natural noise distributions. Additionally, while our evaluation focuses on textual consistency, incorporating user studies would provide complementary insights into perceived trust and explanation reliability. Exploring robustness-aware training objectives and evaluating larger or instruction-tuned LLMs represent promising directions for developing more stable and trustworthy explanation agents in real-world recommender systems.

\section{Conclusion}
\label{sec:conclusion}

We presented RobustExplain, the first systematic evaluation framework for assessing LLM-generated recommendation explanation robustness under user behavior perturbations. Our framework introduces five perturbation types modeling realistic data variations and multi-dimensional robustness metrics capturing semantic, keyword, structural, and length aspects of explanation stability.

Experiments across four LLMs (7B-70B parameters) reveal that current models exhibit moderate sensitivity, with robustness scores averaging around 0.50. Larger models demonstrate measurable stability benefits: LLaMA 3.1-70B achieves 8\% higher robustness than 7-8B models. Notably, robustness remains stable across severity levels with only 1.7\% degradation from mild to severe perturbations, suggesting that LLMs capture high-level preference patterns rather than relying on individual data points. These findings establish important baselines for robust explanation generation and provide guidance on model selection for stability-critical applications.

\bibliographystyle{ACM-Reference-Format}
\bibliography{references}

\end{document}